\documentclass[12pt]{article}

\newenvironment{references}{
  \begin{center} \textsf{REFERENCES} \end{center}
  \begin{list}{}{\topsep=0pt\parsep=0pt\baselineskip=20pt
   \leftmargin=1.5em\itemindent=-\leftmargin}}
  {\end{list}}

\begin{document}

\noindent [Reviewer]\\ 
J. Andrew \textsc{Royle} \\ 
USGS Patuxent Wildlife Research Center \\ 
12100 Beech Forest Road \\ 
Laurel MD 20708 \\ 
301-497-5846 \\ 
aroyle@usgs.gov \\

\noindent [xxxx xxxxx]\\ 
{\Large\sf Mixed Effects Models and Extensions in Ecology with
  R}\footnote{I was asked to review this book for The American
  Statistician in 2010. After I wrote the review, the invitation was
  revoked. This is the review.}

\begin{quotation}\noindent
 Alain F. \textsc{Zuur}, Elena N. \textsc{Ieno}, Neil J. \textsc{Walker}, Anatoly A. \textsc{Saveliev}, and Graham M. \textsc{Smith}.  
New York, NY: Springer, 2009.  ISBN: 978-0-387-87457-9. xxii+574 pp. \$84.95 (H).  
\end{quotation}\vspace{12pt}

\setlength{\baselineskip}{20pt}


This book covers essential material for applied statisticians working
in ecology, natural resources and related fields.  The authors cover
the basics of mixed models, and models for counts: binomial, Poisson,
negative binomial -- the bread and butter of statistical modeling in
ecology -- while emphasizing random effects, variance
heterogeneity, over-dispersion, dependence, and special situations
including zero-truncation and zero-inflation.  The authors achieve 
broad coverage of the focal material, from a classical (likelihood,
frequentist) perspective,
and they do so in a practical
manner and with a writing style that is clear and accessible.  The
result is a book that is easy to read, and I think enormously useful
for practicing statisticians working in ecology and natural
resources. I enjoyed reading this book, and it now resides on my
``within-reach'' shelf containing books that I consult regularly, as I
expect to do with this book.
I have a great resource for locating
specific technical details or context, key references for
a method,  illustrative examples, an ${\bf R}$ package for some 
class of models, or other material related to 
 topics that come up routinely in my interactions
with biologists.

The book is
based on the authors real experience consulting and collaborating with biologists
and that shows in the writing -- clear and not overly-technical
explanations of
topics with attention to how to get things done, 
and what it all means. There are many very interesting data
sets, and the analyses are insightful and comprehensive.  A strength
of the book is that it adopts a framework for analysis based on the
{\bf R} programming language, and it is structured as a kind of ``how
to'' guide, with recipes for each specific analysis. In
writing an applied book of this sort, one of the challenges to
confront is how much ``code'' to put in the book. Too much is
distracting, and too little makes the book less useful to
less-advanced readers.  
The authors achieve the perfect balance for the level of their book.

The structure of the book is 13 chapters covering the core
methodological content, which is tied together nicely with meaningful
examples. 
At chapter 14, the format changes slightly and there is a
sequence of 10 chapters (Chs. 14-23) where the real world is
confronted in the form of more in-depth case studies.  While specific
authors are not acknowledged for the first 13 chapters, the case study
chapters are coauthored by specific individuals, including the authors
of the book but including also their collaborators.  There is one
appendix covering regression and other basic applied concepts that are
assumed to be prior knowledge.

The organization and topical coverage of the first 13 chapters is a
great
strength of the book.
This is essential material
that is not often covered in such detail, and in such 
a coherent, integrated manner.
The organization is ideal for a workshop or class
setting. 
Chapter 2 deals with linear Regression,
or rather its ``limitations'' -- diagnosing violations of 
assumptions. Following this are chapters on what to do about it.
Chapter 3 considers additive models, i.e., generalized additive
models -- ``GAMS'' -- a methodological framework which the authors are
very fond of, and is used throughout the book.
While there is a lot of good material on GAMs (what they are, how to fit them, and actual
applications), my sense is that GAMs are somewhat over-emphasized
relative to how much you need them in practice.
That said, it is useful
to have exposure to this material and 
coverage of GAMs to such an extent is relatively uncommon in ecological texts.
Chapter 4 deals with
modeling variance heterogeneity and weighted least-squares -- kind of a ``pre-mixed models'' chapter.  
This is good material to segue into mixed modeling and it provides useful methodological and
conceptual context. In chapter 4 they
introduce ``the 10-step protocol'' (p. 90) for model selection in
mixed
models. On my first read I was not
sure exactly what the point of the protocol was -- it appears rather
abruptly
and with not enough context, motivation or justification.
The authors really push ``the protocol'' throughout the book.  I
haven't decided if I really like it, but I suppose it 
gives some practitioners a decision tree that they can apply to any
problem. 
Chapter 5 has good material on mixed models, random intercepts and slopes models,
induced correlation structures,
model selection, REML and related topics.
Chapter 6 covers what to do about ``dependence.'' This is great
material that is not covered enough in applied statistics classes.
Chapter 7 expands on that topic in the context of spatial dependence.
Chapter 8 is kind of an introduction to modeling non-normal data,
covering specific distributions (Poisson, neg binomial, Gamma, etc.)
Chapter 9 is focused on Poisson and negative binomial GLMs and GAMs, 
and Chapter 10 basically the same but for binomial data.
Chapter 11 covers zero-inflated and zero-truncated models for count
data -- material that is very useful, and under-emphasized in most texts.
Chapter 12 extends mixed modeling concepts to count data with the use
of generalized estimating equations (GEEs). 
Chapter 13 contains a very brief treatment of generalized linear/additive mixed
models (GLMMs and GAMMs).


One criticism I have is primarily a philosophical one which can best
be summed up by a remark that I recently read in a book by K\'{e}ry
(2010) (a book largely having to do with Bayesian analysis of mixed
models). The following passage really resonated with me:
 ``In statistics classes at university, many ecologists have only
seen a sad caricature of statistics. We were taught to think in terms
of a decision tree for black-box procedures. The tree started with a
question like ``Are the data normally distributed?'' and its terminal
branches prescribed a t-test or a Kruskal-Wallis test or an analysis
of covariance with homogeneous slopes (or else we were in deep
trouble). And then, a p-value popped up somewhere, and if it was
$<0.05$, life was good.''  At certain times, elements of this book are 
a little bit
too close to this ``sad caricature'' of statistics. e.g., the 10 step
protocol is a version of this ``decision tree'', with just a little
too much focus on procedure with p-values as the objective.  But,
fortunately, the authors are not overly dogmatic about these things
(the protocol, and model selection) and I think the strengths of the
organization and coverage outweigh my philosophical misgivings.

In terms of topical coverage, I think there are two important
methodological elements missing from the book.  First is that Bayesian
analysis is not covered in a meaningful way and, perhaps
coincidentally, there is limited coverage of GLMMs. This because, as
the authors acknowledge, such methods are ``...on the frontier of
statistical research.''  The reader is left thinking that GLMMs must
be intractable when, in fact, the mechanics of Bayesian inference
under such models doesn't really require any special considerations.
I believe that most ecologists have quite an interest in GLMMs and,
over the last several years, they have become routinely used in
ecology, and analyses based on such models appear regularly in
ecological journals and many recent books. There are now several
convenient and accessible computing platforms for Bayesian analysis.
Another element missing from this book is that the authors don't cover
certain classes of models that are somewhat specialized to ecology such as
distance sampling, or capture-recapture for estimating abundance,
density, survival, recruitment and movement rates.  Population level
studies represent an important segment of the market, and there is
virtually nothing in this book to help the ecologist down that road.
While such topics are widely covered in many other books (e.g.,
Williams et al. 2003), their absence here presents a view that
ecologists can solve all of their problems using various flavors of
regression.

In the grand scheme of things, these are probably minor omissions
necessitated by the conceptual and methodological focus adopted by the
authors.  And, this is 
compensated for by the comprehensive and practical coverage of so much
 ``gotta-know'' material on the core methodological content. This is
 why I think 
this book is one that applied statisticians and practitioners
should have. Put it on your ``within-reach'' shelf.


\begin{flushright}\def\baselinestretch{1}
J. Andrew \textsc{Royle} \\ 
\emph{USGS Patuxent Wildlife Research Center}
\end{flushright}




\end{document}